\documentclass{ws-ijmpc}

\usepackage{color}
\usepackage{graphicx}
\usepackage{float}
\usepackage[caption=false]{subfig}
\begin{document}

\markboth{M.R. Parsa, A. Pachalieva and A.J. Wagner}
{Validity of the Molecular-Dynamics-Lattice-Gas Global Equilibrium Distribution Function}

\catchline{}{}{}{}{}

\title{Validity of the Molecular-Dynamics-Lattice-Gas Global Equilibrium Distribution Function
}

\author{M. Reza Parsa
}

\address{Program in Materials and Nanotechnology, North Dakota State University, Fargo, ND 58108\\
Department of Physics, North Dakota State University, Fargo, ND 58108
mohammad.parsa@ndsu.edu}

\author{Aleksandra Pachalieva}

\address{Department of Mechanical Engineering, Technical University of Munich,\\
85748 Garching, Germany\\
aleksandra.pachalieva@tum.de}

\author{Alexander J. Wagner}

\address{Department of Physics, North Dakota State University, Fargo, ND 58108\\
alexander.wagner@ndsu.edu}

\maketitle

\begin{history}
\received{Day Month Year}
\revised{Day Month Year}
\end{history}

\begin{abstract}
	The MDLG method establishes a direct link between a lattice-gas method and the coarse-graining of a Molecular Dynamics approach. Due to its connection to Molecular Dynamics, the MDLG rigorously recovers the hydrodynamics and allows to validate the behavior of the lattice-gas or lattice-Boltzmann methods directly without using the standard kinetic theory approach. In this paper, we show that the analytical definition of the equilibrium distribution function remains valid even for very high volume fractions.
\end{abstract}

\ccode{PACS Nos.: 11.25.Hf, 123.1K}


\section{Introduction}

The lattice gas automata were introduced by Frisch, Hasslacher and Pomeau\cite{Frisch.1986} in 1986. These methods describe the presence of a particle using Boolean states and thus exhibit perfect collisions. However, the lattice gas methods suffer from statistical noise and their collision rules can have very complex mathematical representation\cite{WolfGladrow.2000}. Later on, Boghosian et al.\cite{Boghosian.1997} introduced the integer lattice gases, where one can control the level of fluctuations, while maintaining the exact conservation laws and having unconditional stability. In the pursuit of understanding better the fluctuating systems, Blommel et al.\cite{Blommel.2018} constructed a new integer lattice gas with Monte Carlo collision operator. 

The Molecular-Dynamics-Lattice-Gas (MDLG) method\cite{Parsa.2017,Parsa.2018} establishes a direct link between a lattice gas method and the coarse-graining of a molecular dynamics (MD) simulation. After comparing the equilibrium properties of the MDLG method to the lattice Boltzmann equilibrium, Parsa et al.\cite{Parsa.2017} found that for any dilute gases for coarse-graining lattice spacing $\Delta x$ exists a coarse-grained time step $\Delta t$ such that the MDLG equilibria resembles the lattice Boltzmann method. However, an open question remained about the range of validity for the predicted analytical solution of the equilibrium distribution function.

 \textcolor{black}{The novelty of the current publication is to investigate the behavior of the MDLG method for varying volume fractions of the underlying MD simulation. Such a test was not performed in the initial publication by Parsa et al.\cite{Parsa.2017}.} This analysis shows that the analytical MDLG equilibrium function remains valid even for very high volume fractions. 

The rest of the paper is summarized as follows: In Section \ref{sec:mdlg}, we elaborate upon the main components of the MDLG method. Our findings on how the equilibrium distribution function behaves with varying nominal volume fraction are presented in Section \ref{sec:fi_analysis}. Finally, some concluding remarks and future discussions are mentioned in Section \ref{sec:conclusions}.

\section{Molecular-Dynamics-Lattice-Gas Method}
\label{sec:mdlg}

\begin{figure}[h]
	\centering
	\subfloat[$a^2 = 0$]{\includegraphics[scale=0.14]{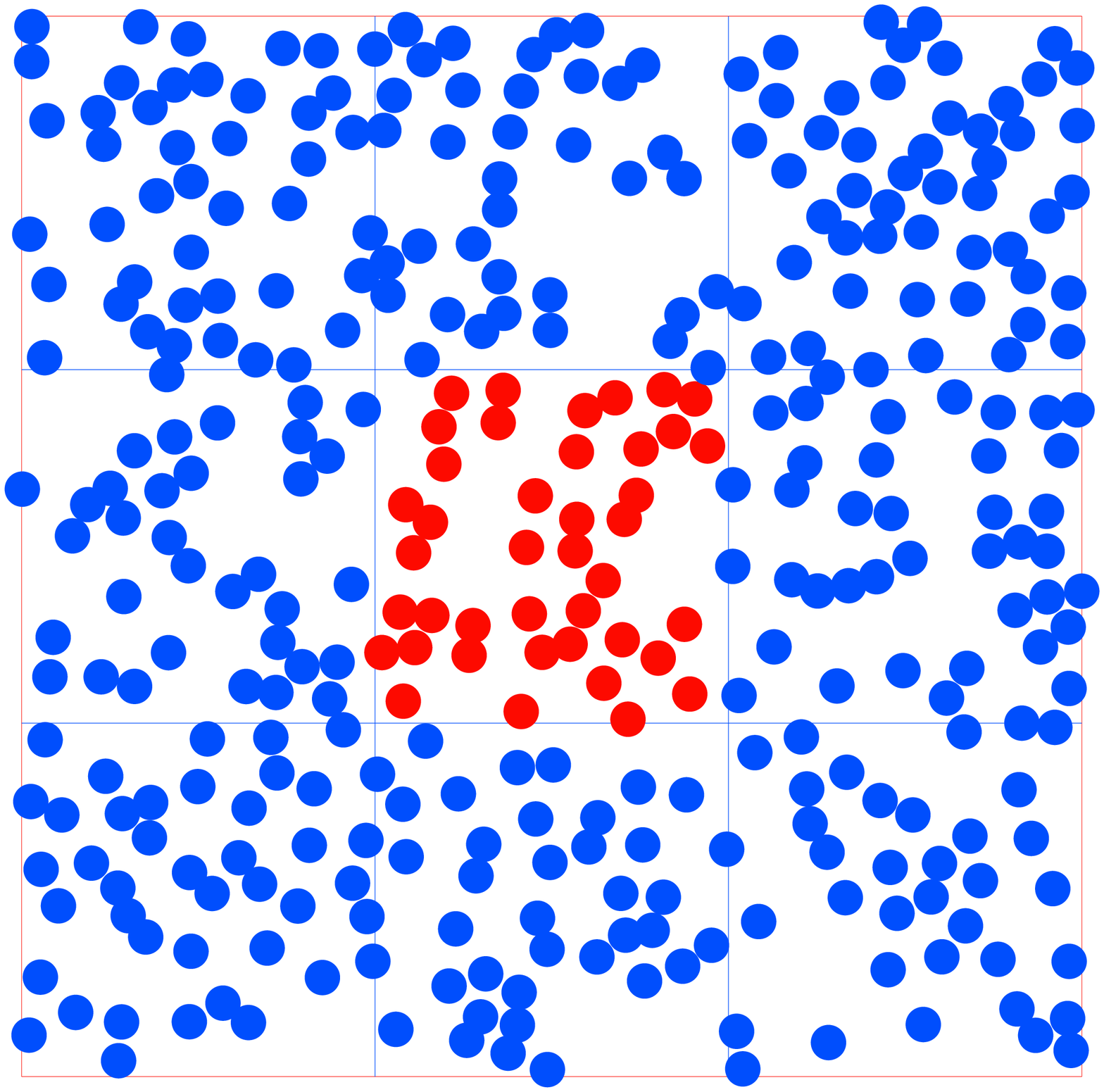}} \hspace{0.1cm}
	\subfloat[$a^2 = 0.003$]{\includegraphics[scale=0.14]{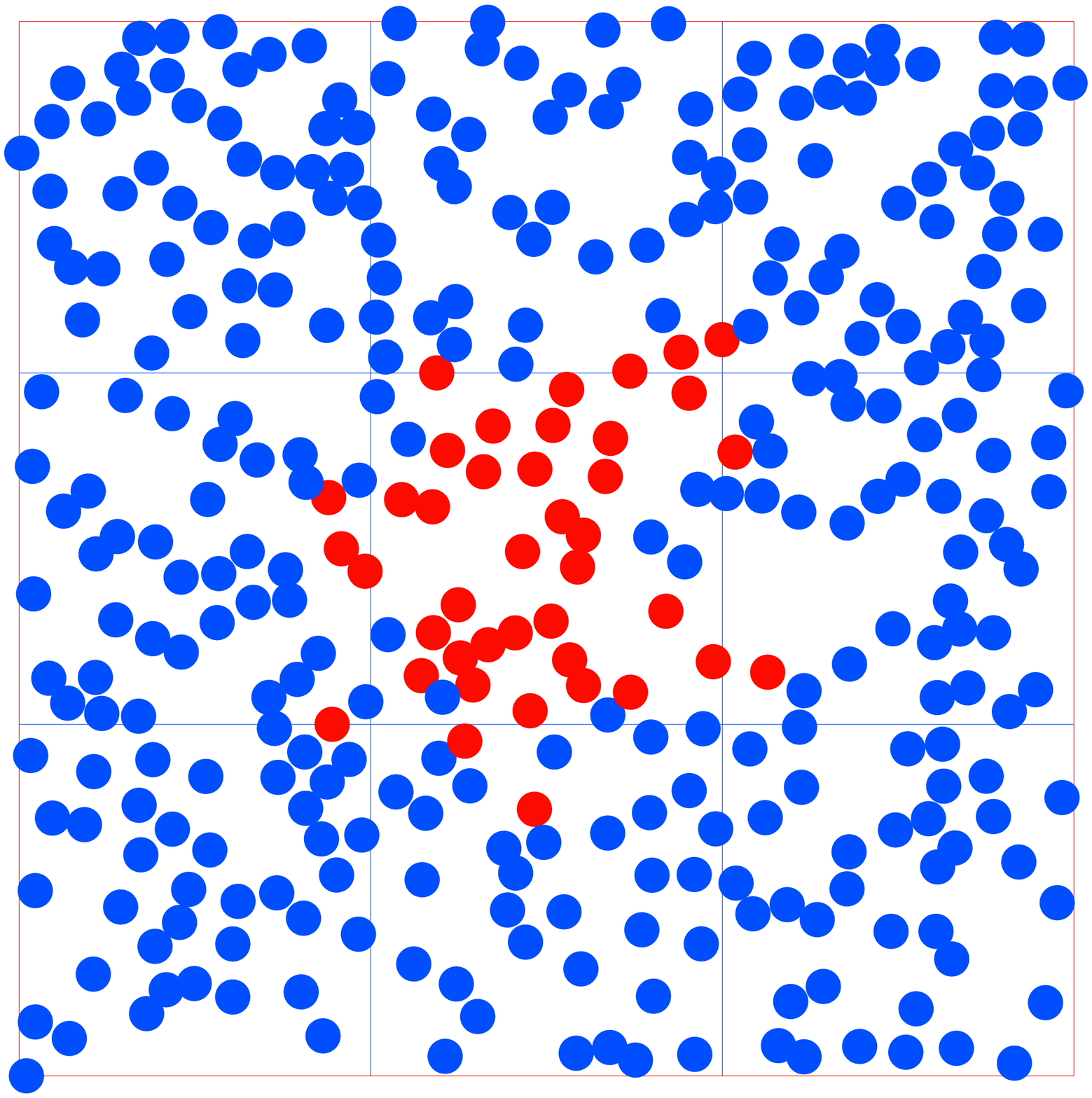}} \hspace{0.1cm}
	\subfloat[$a^2 = 0.010$]{\includegraphics[scale=0.14]{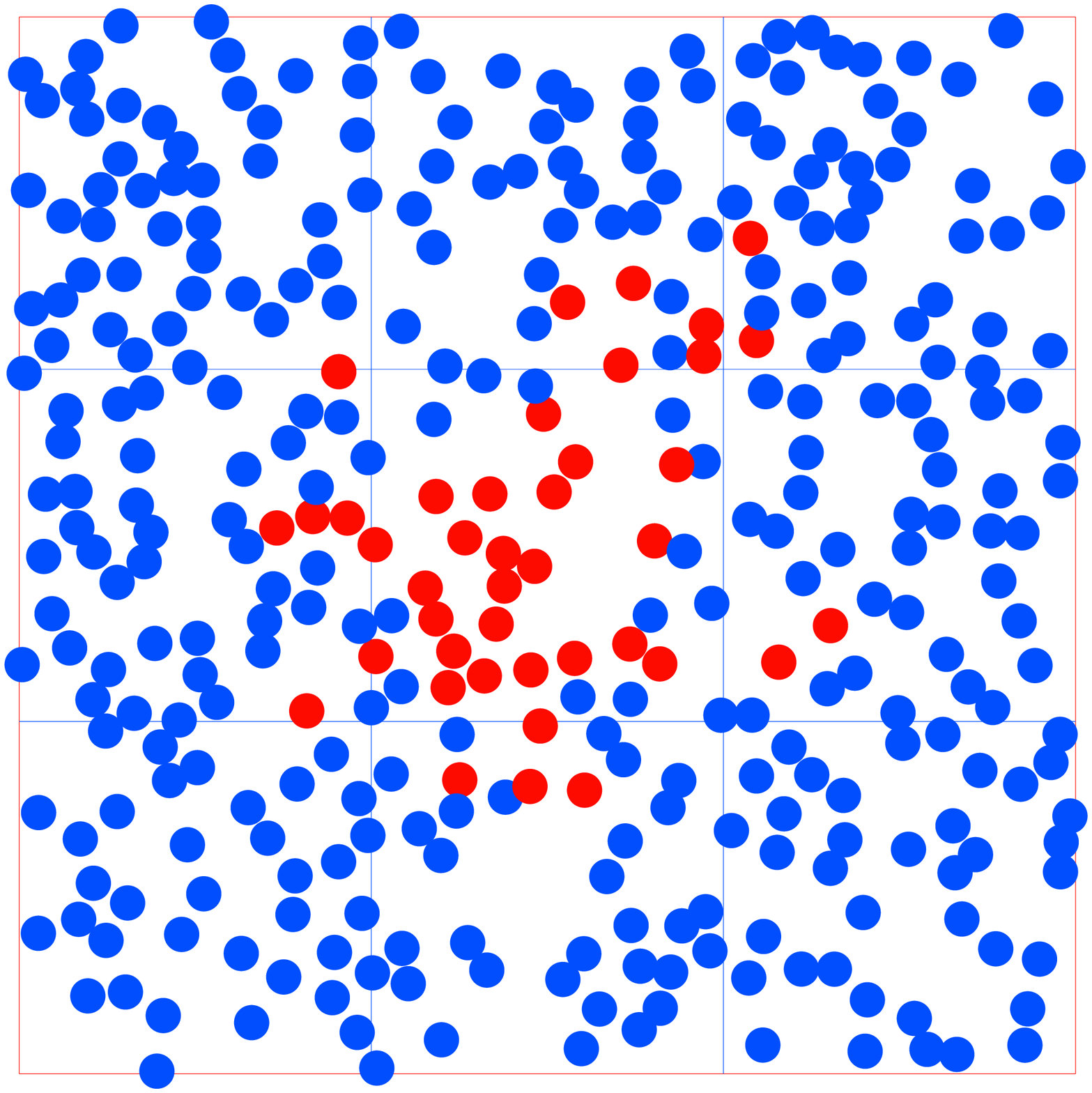}} \hspace{0.1cm}
	\subfloat[$a^2 = 0.019$]{\includegraphics[scale=0.14]{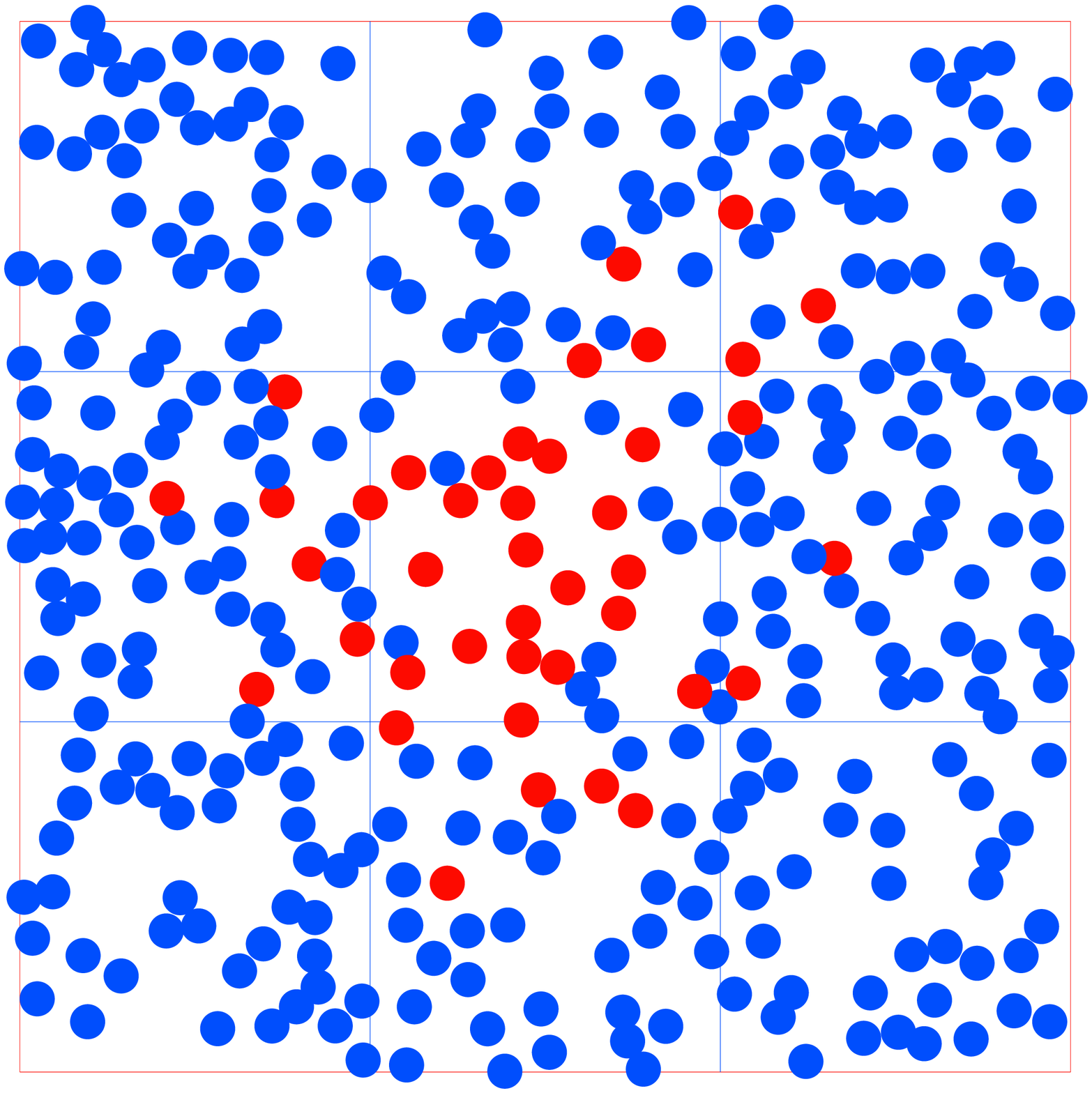}}
	\vspace*{8pt}
	\caption{\textcolor{black}{(Color online) Representation of the MDLG algorithm: a lattice (blue line) is overlaid onto the MD simulation. The position of each particle is tracked depending on an a priori chosen time ($\Delta t$) and space ($\Delta x$) discretization for the MDLG algorithm. The particles in the central lattice are colored in red to allow the reader to track their movement, however, each particle in the MDLG method is tracked and its occupation has been considered at each time step. The results are obtained from a Md simulation with volume fraction of $\phi=0.0785$, $\kappa_B T = 50$ in Lennard-Jones (LJ) units, $\lambda = 0.5$, $\Delta x = 50$ and $\Delta t = 0.0001$. The shown time frames correspond to (a) $t = 0$, (b) $t = 0.3$, (c) $t = 0.6$, (d) $t = 0.9$ in LJ units.}}
	\label{fig:mdlg_definition}
\end{figure}

The MDLG method was originally introduced by Parsa et al.\cite{Parsa.2017}. It is based on an underlying MD simulation, where we track the migration of the particles to imposed lattice positions with displacement $v_i$ after a time step $\Delta t$, as shown in Fig. \ref{fig:mdlg_definition}.
\vspace*{-0.2cm}
This gives the integer lattice gas occupation number
\begin{equation}
	n_i(x,t) = \sum_{j}\Delta_x[x_j(t)]\Delta_{x-v_i}[x_j(t-\Delta t)],
\end{equation}
where $\Delta_x[x_j(t)]=1$, if particle $x$ is in lattice cell $x$ at time $t$, and  $\Delta_x[x_j(t)]=0$, otherwise. 

The MDLG evolution equation then takes the form of a lattice gas
\begin{equation}
n_i(x+v_i, t+\Delta t) = n_i(x,t) + \Xi_i,
\label{eq:mdlg_evolition}
\end{equation}
where the collision operator $\Xi_i$ is defined as
\begin{equation}
\Xi_i = n_i(x+v_i, t+\Delta t) - n_i(x,t).
\label{eq:mdlg_collision}
\end{equation}
Since this analysis is based on a MD simulation, the resulting lattice gas model rigorously conserves the hydrodynamic properties of the system, up to the coarse-graining approximation.
We define a Boltzmann average of the MDLG as
\begin{equation}
	f_i = \langle n_i\rangle_{neq}
	\label{eq:fi_from_ni}
\end{equation}
and by taking an non-equilibrium ensemble average of Eq. (\ref{eq:mdlg_evolition}), we obtain the MDLB evolution equation: 
\begin{equation}
	f_i(x+v_i, t+\Delta t) = f_i(x,t) + \Omega_i \qquad\mbox{with} \qquad\Omega_i = \langle\Xi_i\rangle_{neq}.
	\label{eq:fi_mdlg_evolition}
\end{equation}
The global equilibrium distribution function can be numerically approximated by averaging the lattice gas densities $n_i$ over the whole lattice and all iterations for an equilibrium system:
\begin{equation}
f_i^{eq} = \langle n_i \rangle_{eq}= \sum\langle\Delta_x[x_j(t)]\Delta_{x-v_i}[x_j(t-\Delta t)]\rangle.
\label{eq:analytical_fi_eq_1}
\end{equation}
Under the assumption that for an ideal gas the displacements $\delta x$ of the particles are independent and that their probability distribution is a Gaussian with variance given by the mean-squared displacement $\langle(\delta x)^2\rangle$ the equilibrium function can be analytically predicted to be
\begin{equation}
	\frac{f_i^{eq}}{\rho^{eq}} = \prod_{\alpha=1}^{d}f_{i, \alpha}^{eq},
	\label{eq:analytical_fi_eq_2}
\end{equation}
where
\begin{align}
  f_{i,\alpha}^{eq} &= N\Bigg( e^{-\frac{(u_{i,\alpha}-1)^2}{2a^2}} - 2e^{-\frac{u^2_{i, \alpha}}{2a^2}} + e^{-\frac{(u_{i,\alpha}+1)^2}{2a^2}}\Bigg) \nonumber\\
  &+ \frac{u_{i, \alpha}-1}{2}\left[\text{erf}\left(\frac{u_{i,\alpha}-1}{\sqrt{2}a}\right) - \text{erf}\left(\frac{u_{i,\alpha}}{\sqrt{2}a}\right)\right]
\label{eq:analytical_fi_eq_3}
\\ &+ \frac{u_{i, \alpha}+1}{2}\left[\text{erf}\left(\frac{u_{i,\alpha}+1}{\sqrt{2}a}\right) - \text{erf}\left(\frac{u_{i,\alpha}}{\sqrt{2}a}\right)\right]
\nonumber
\end{align}
and
\begin{align}
	a^2 = \frac{\langle (\delta x_\alpha)^2 \rangle}{(\Delta x)^2},\qquad
	N = \frac{a}{\sqrt{2\pi}},\qquad
	u_{i,\alpha} = v_{i\alpha} - u_\alpha.
	\label{eq:analytical_fi_a2}
\end{align}
where $\Delta x$ is the lattice size. \textcolor{black}{This is the main result from the publication by Parsa et al.\cite{Parsa.2017} and the derivation can be found there. The weights obtained by the MDLB method resemble to high extend the standard D2Q9 weights ($w_0=4/9, w_{1-4}=1/9, w_{5-8}=1/36$) for a specific time and space discretization with $a^2=1/6$. The weights of the MDLB method for $a^2=1/6$ obtained from Eq.(\ref{eq:analytical_fi_eq_2}) are
	\begin{align}
		w_0 &= 0.45721,\\
		w_{1-4} &= 0.10883,\\
		w_{5-8} &= 0.025907,
	\end{align}
where sum over all the D2Q9 velocity weights is slightly lower than 1.0 ($\approx$0.996158) because the MDLB does not impose restrictions on the number of velocities and thus, higher velocities also have small contributions to the total sum. We can make predictions for the form of higher order lattice velocity sets like the one published in \cite{Shan.2010,Chikatamarla.2009,Atif.2018}. However, this is out of the scope of the current publication. Due to the lack of a velocity set restriction, the MDLB models also do not have a lattice velocity restriction typical for the LBM methods. The equilibrium distribution function given in Eq.(\ref{eq:analytical_fi_eq_2}) does not resemble any of the already published formulations of the equilibrium distribution function because it is not explicitly restricted to a specific velocity set. As a comparison, please, refer to the MCLG model published by Blommel et al.\cite{Blommel.2018}, where an apriori restriction of the number of velocities is made and the equilibrium distribution function obtained by the authors recovers the entropic formulation of the equilibrium distribution function given by Ansumali et al\cite{Ansumali.2003}.}
\\
For ideal gas systems the mean-square displacement is given in relation to the velocity correlation function
\begin{align}
	\langle(\delta x)^2\rangle = 2\int_{0}^{t} dt'(t-t')\langle v_\alpha(t')v_\alpha(0)\rangle,
	\label{eq:msd_analytic}
\end{align}
which is well approximated by an exponential decay given by
\begin{align}
\langle v_\alpha(t')v_\alpha(0)\rangle = k_BT\exp{\left(-\frac{t}{\lambda}\right)},
\label{eq:velocity_corr_fn}
\end{align}
with $k_B$ being the Boltzmann constant, $T$ the temperature in Lennard-Jones (LJ) units, and $\lambda$ the exponential decay constant\cite{Green.1952,Kubo.1957,Weitz.1989}. Now, we can express the mean-square displacement as a function of a single free parameter $\lambda$:
\begin{align}
	\langle(\delta x)^2\rangle = 2\kappa_B T \lambda^2\left(\exp{\Big(-\frac{t}{\lambda}\Big)}+\frac{t}{\lambda}-1\right). 
	\label{eq:msd_predicted}
\end{align}
that is depicted in Fig. \ref{fig:mean_sq_displ}. 

The assumption of uncorrelated displacements, which was used for the prediction of the equilibrium distribution, is likely only valid when the system can be approximated as an ideal gas.  Our aim in this paper is, therefore, to find the validity range of this assumption by comparing the measured equilibrium distribution and the theoretical prediction for different nominal volume fractions.


\section{MDLG Equilibrium Distribution Function for Different Volume Fractions}
\label{sec:fi_analysis}
\begin{figure}[h]
    \centering
    \subfloat[$\phi=0.0078$]{\includegraphics[scale=0.14]{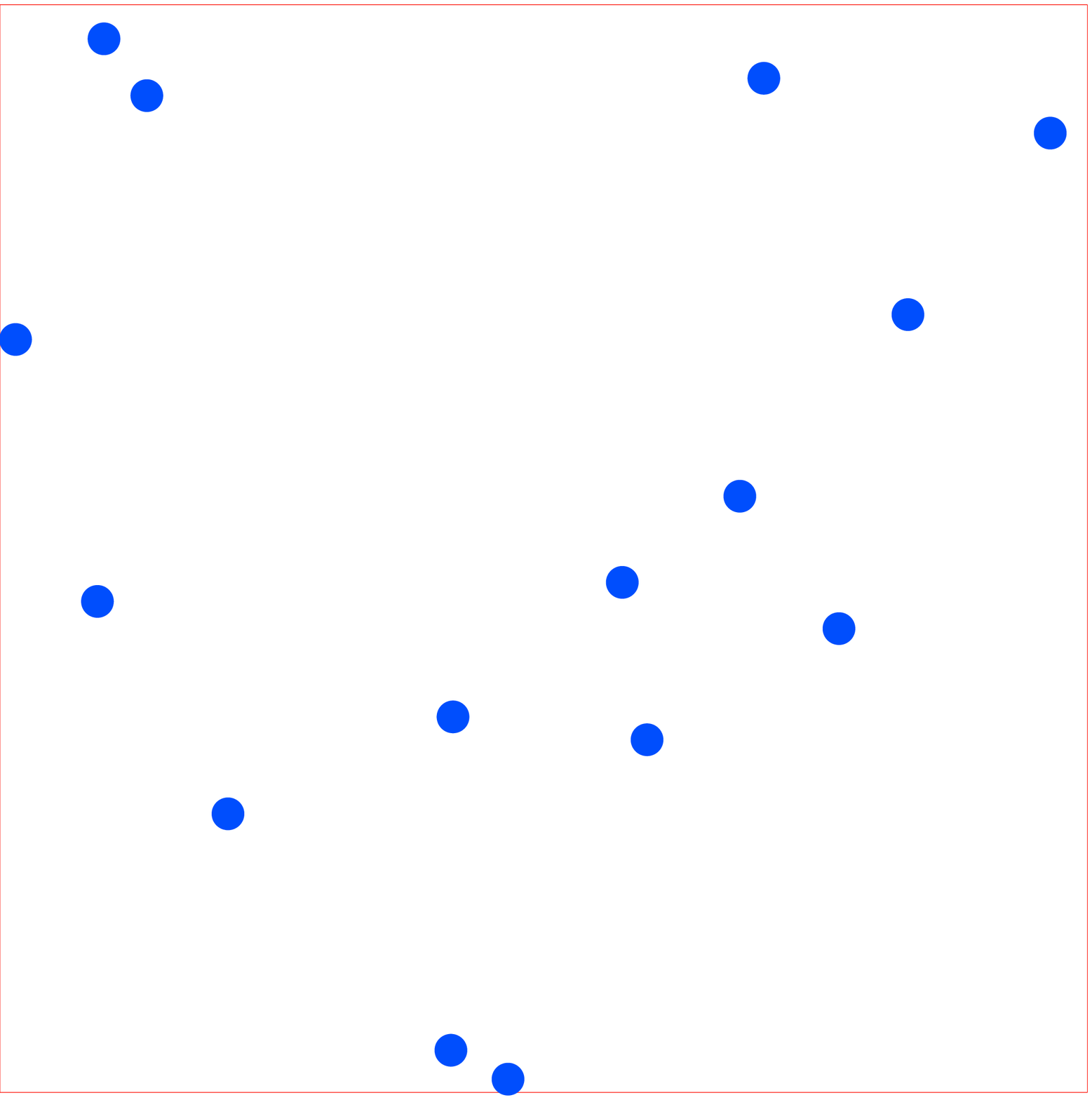}}\hspace{0.1cm}
    \subfloat[$\phi=0.0785$]{\includegraphics[scale=0.14]{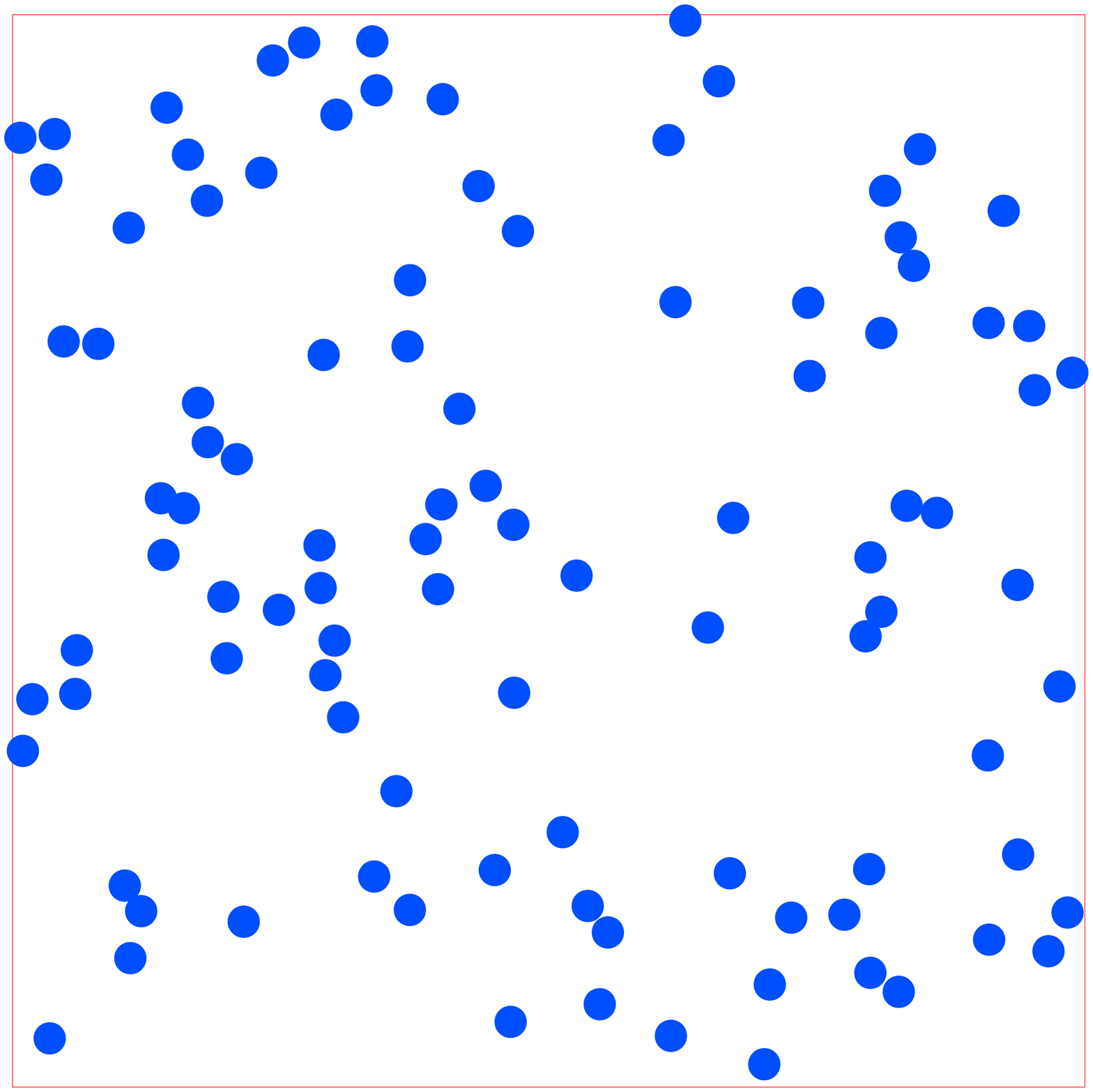}}\hspace{0.1cm}
    \subfloat[$\phi=0.1962$]{\includegraphics[scale=0.14]{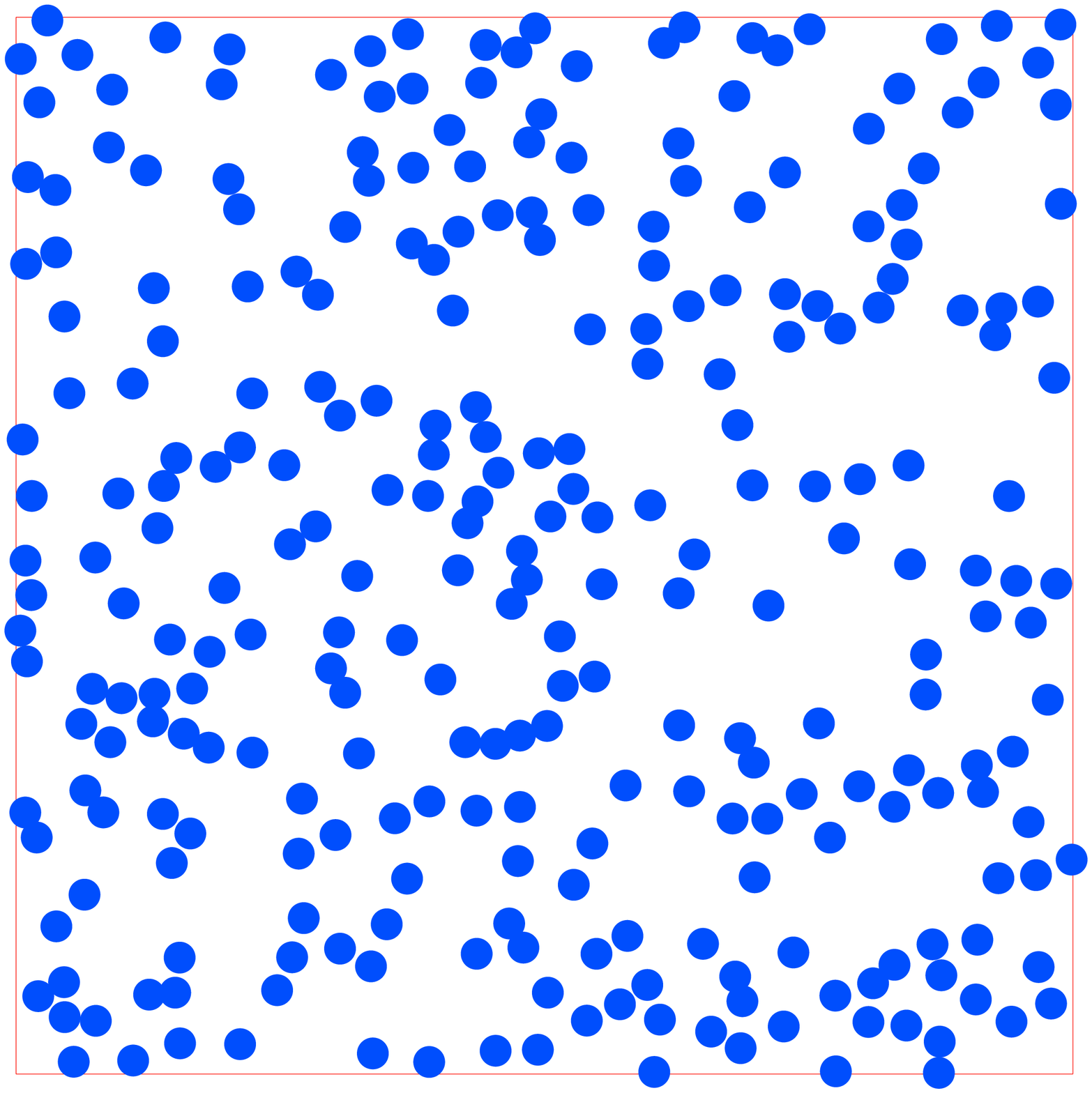}}\hspace{0.1cm}
    \subfloat[$\phi=0.8722$]{\includegraphics[scale=0.14]{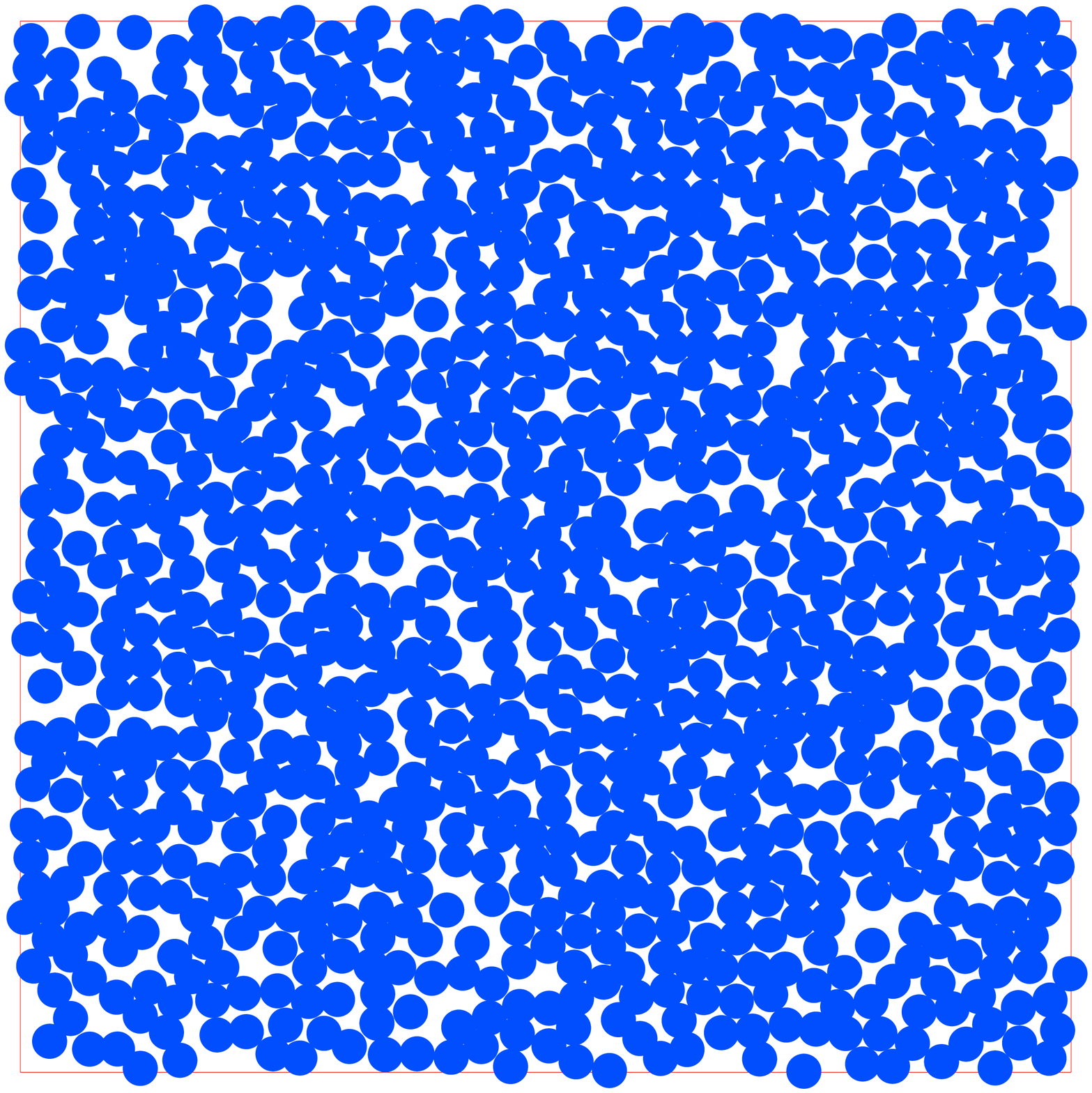}}
    \vspace*{8pt}
\caption{(Color online) Visual representation and comparison between the volume fractions used for the underlying MD simulations. The LJ particles are represented by circles with radius $r=\sigma$ with $\sigma$ being the distance at which the inter particle potential is zero.}
\label{fig:nominal_volume_fraction}
\end{figure}

We have chosen four different setups with varying volume fractions for our underlying MD simulations with standard Lennard-Jones (LJ) interaction potential defined as
\begin{equation}
    V_{LJ}=4\varepsilon\left[\left(\frac{\sigma}{r}\right)^{12}-\left(\frac{\sigma}{r}\right)^{6}\right]
\end{equation}
where $\varepsilon$ is the depth of the potential well, $\sigma$ the distance where the LJ potential is zero and $r$ the distance between particles.
The MD simulations were performed using the open-source LAMMPS framework. We vary the volume fraction $\phi$ of the MD simulations from $\phi=0.0078$ to $\phi=0.8722$ (where the low value corresponds to the density employed in \cite{Parsa.2017}). \textcolor{black}{The velocity has been fixed to $u=0$, to allow for a deliberative analysis of the MDLG method in respect only to the volume fraction. For details, how the system behaves for different velocities, please, refer to \cite{Parsa.2017}}. The volume fraction $\phi$ is defined for circular LJ particles with radius $r=\sigma$ with $\sigma$ being the distance at which the inter particle potential goes to zero. Even though, the simulated volume fraction is above the maximum package density for hard spheres ($\phi=0.7405$), we still observe diffusion due to the high temperature of the system (50 LJ units). When we increase the volume fraction even more, the system goes into a solid state and the dynamics slows down considerably. A visual representation of the variation of the used volume fractions can be seen in Fig.  \ref{fig:nominal_volume_fraction}(a)-(d).
\vspace*{-0.2cm}

All the simulations were initialized with homogeneously distributed particles. The total number of measured iterations for each simulation setup is $2,000,000$ with a time step of $0.0001\tau$, which corresponds to a timescale of $\tau=200$. As in the previous study\cite{Parsa.2017}, a sufficient number of initial iterations, ($10^6, 10^6, 10^6, 10^7$, respectively), were discarded from the sampling process to ensure that the system has reached an equilibrium state before the probing. 
\begin{figure}[b]
	\centering
	\subfloat{\includegraphics[scale=0.45]{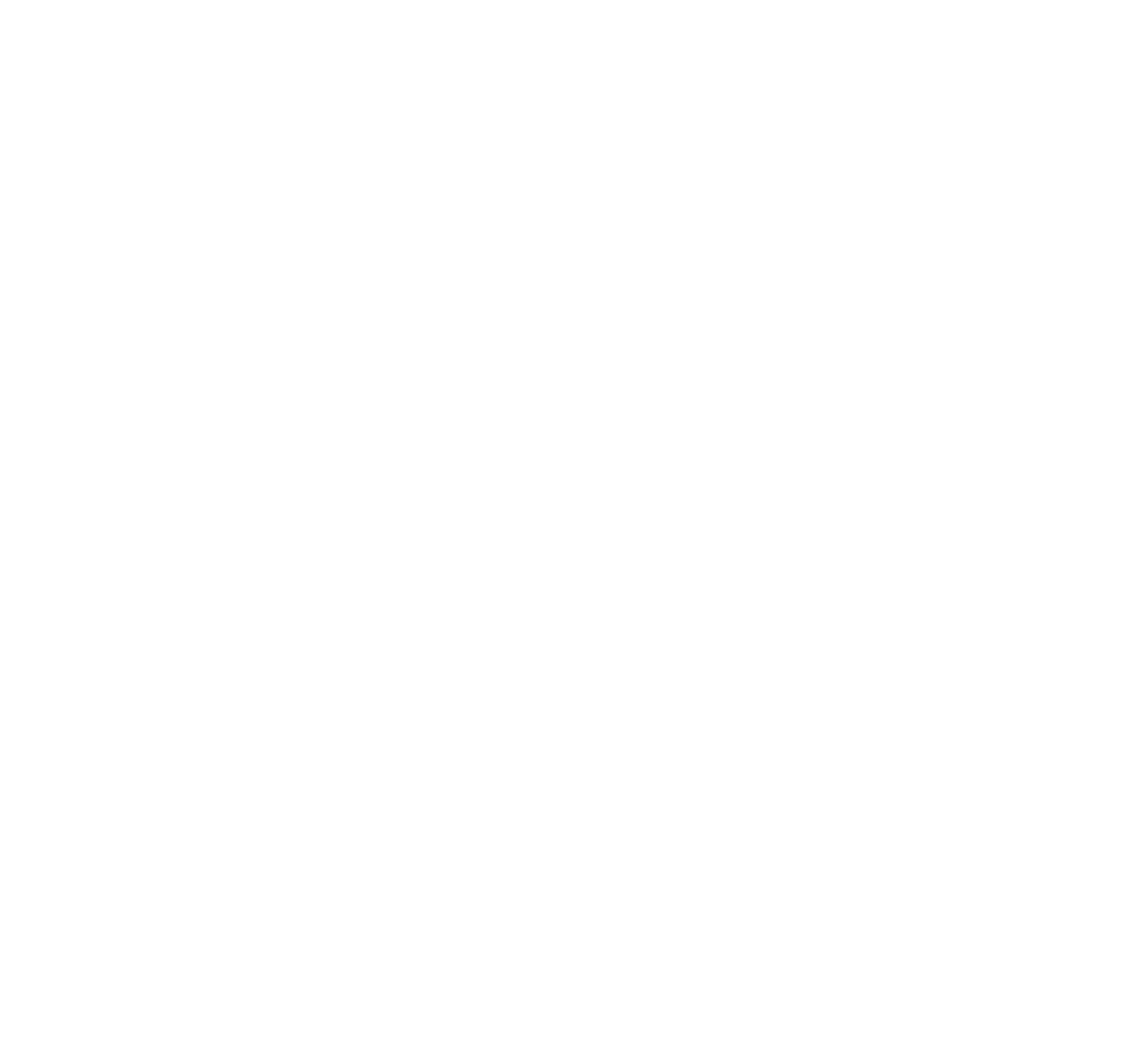}}
	\vspace*{8pt}
	\caption{(Color online) Measured mean-square displacement from the MD simulation data (symbols) for different volume fractions compared to the corresponding predicted values (dashed line).}
	\label{fig:mean_sq_displ}
\end{figure}

After fitting the exponential constant $\lambda$ of the velocity correlation function in Eq. (\ref{eq:velocity_corr_fn}) from the MD data, we can use the relation between the velocity correlation function and the mean-square displacement to obtain an analytical representation of the mean-square displacement as given in Eq. (\ref{eq:msd_analytic}) for each of the nominal volume fractions (from $\phi=0.0078$ to $\phi=0.8722$). In Fig. \ref{fig:mean_sq_displ}, we compare the predicted mean-square displacement from Eq. (\ref{eq:msd_analytic}) and the measured value from the underlying MD simulation data. There is an excellent agreement between the theoretical mean-square displacement and the measured one based on the velocity correlation exponential fit for volume fractions from $\phi=0.0078$ to $\phi=0.1962$. Those volume fractions describe, in general, more gassy systems. For denser systems with $\phi=0.8722$, the measured mean-square displacement deviate from the predicted one. As shown in Fig. \ref{fig:nominal_volume_fraction}(d), a system with $\phi=0.8722$ is rather in a state close to a melted solid under high pressure.

\begin{figure}[t]
    \centering
    \subfloat[$\phi=0.0078$]{\includegraphics[scale=0.40]{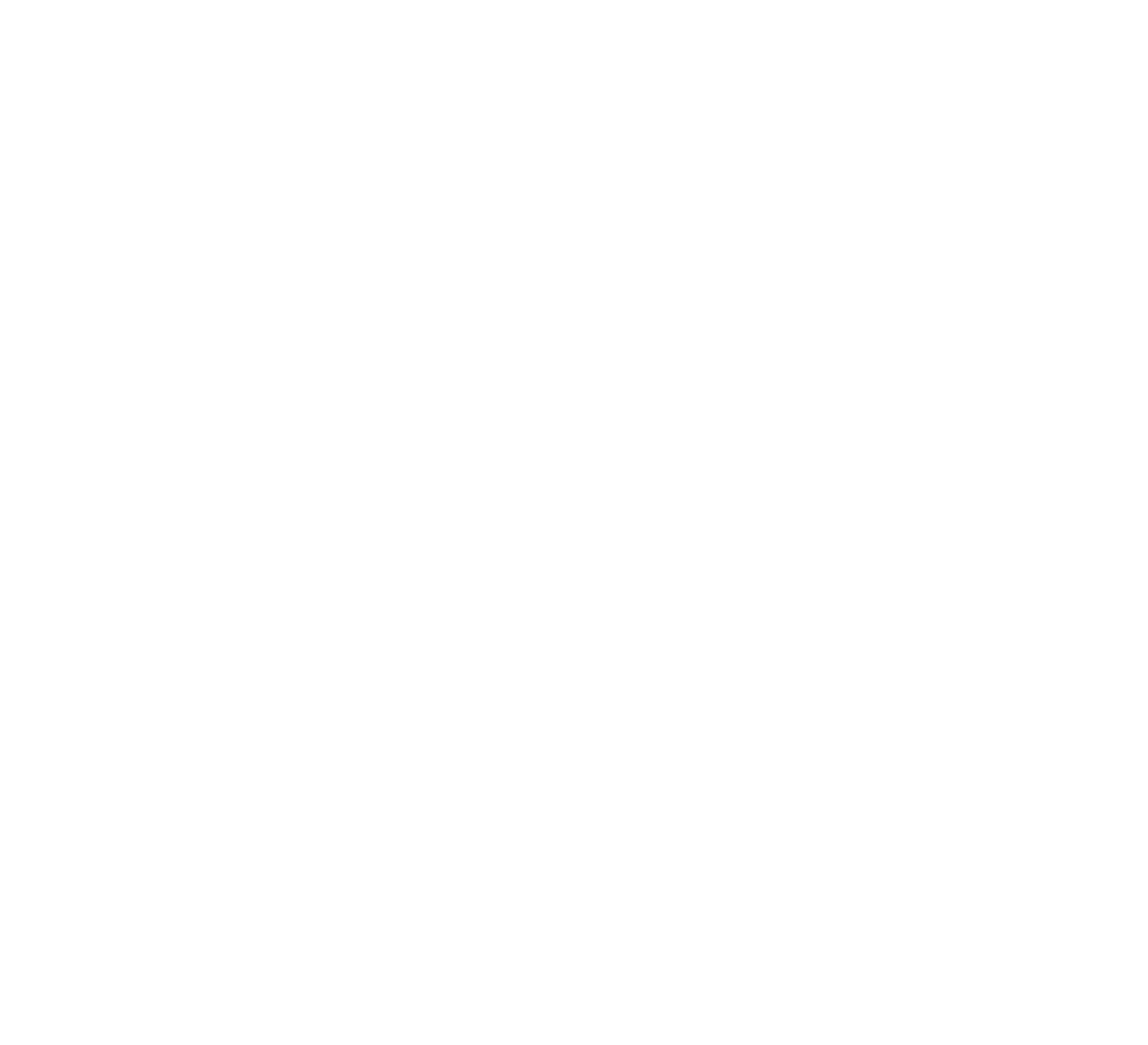}}
    \subfloat[$\phi=0.1962$]{\includegraphics[scale=0.40]{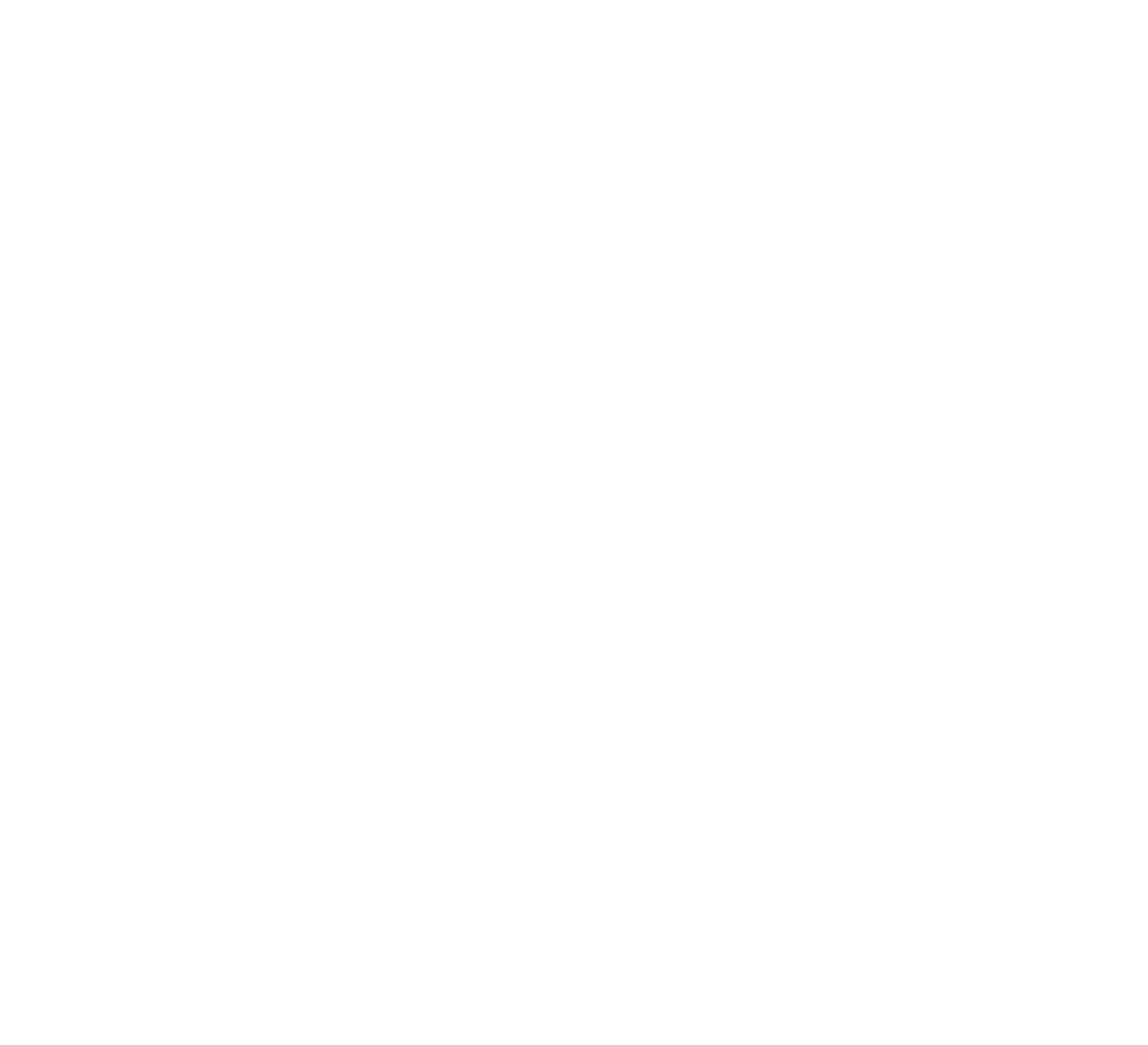}}
    \vspace*{8pt}
    \caption{(Color online) Comparison between the measured equilibrium distribution function $f_i^{eq}$ from an underlying MD simulation and the analytical solution given in Eq. (\ref{eq:analytical_fi_eq_2}) for different volume fractions from $\phi=0.0078$ to $\phi=0.1962$. The results for volume fraction $\phi=0.0785$ has been already published in Parsa et al.\protect\cite{Parsa.2017}, where we observed the same behavior.}
\label{fig:eq_distr_fn1}
\end{figure}

\begin{figure}[h]
    \centering
    \subfloat[Theoretical $\langle(\delta x)^2\rangle$, $\phi=0.8722$ ]{\includegraphics[scale=0.40]{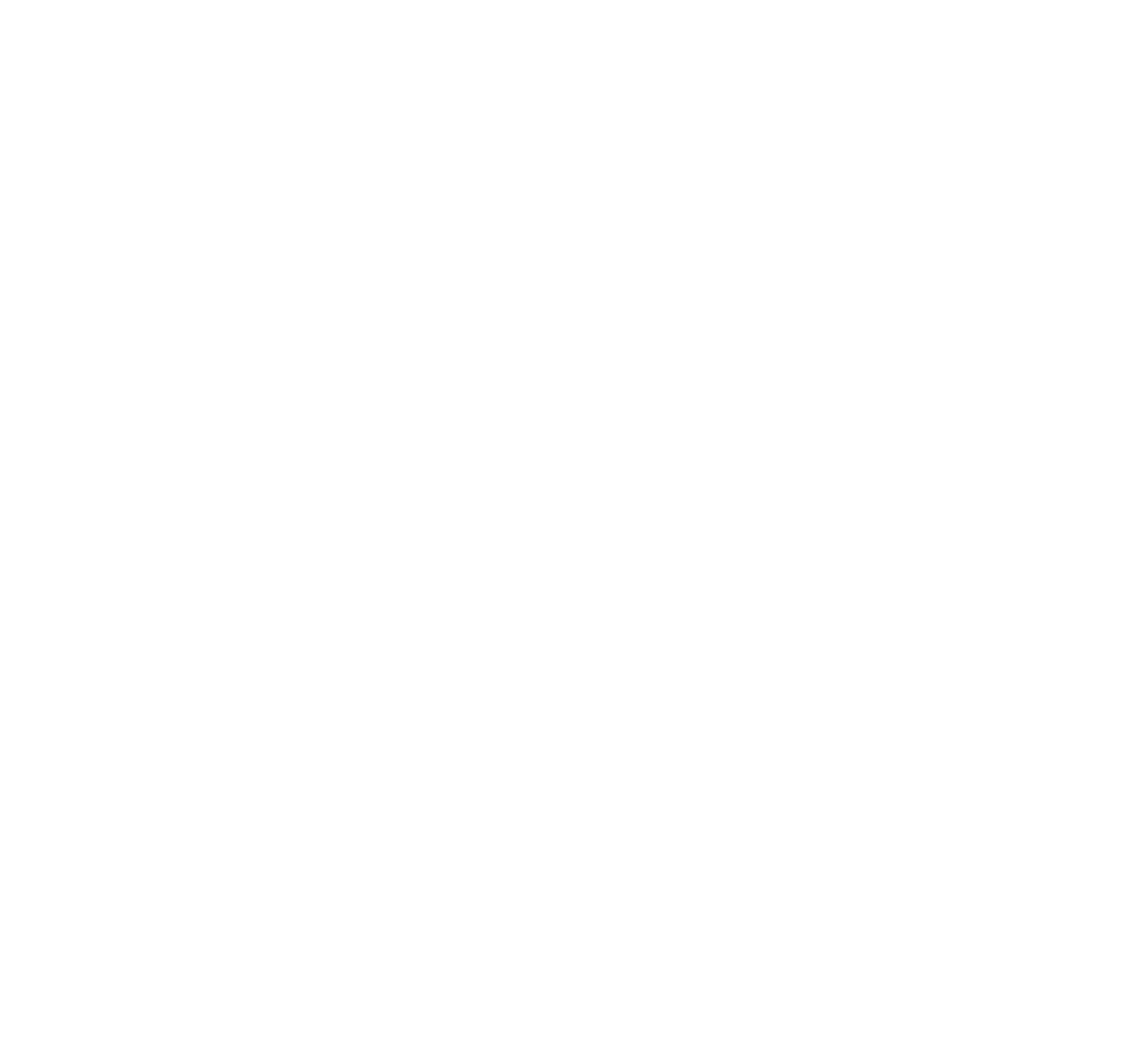}}
    \subfloat[Measured $\langle(\delta x)^2\rangle$, $\phi=0.8722$]{\includegraphics[scale=0.40]{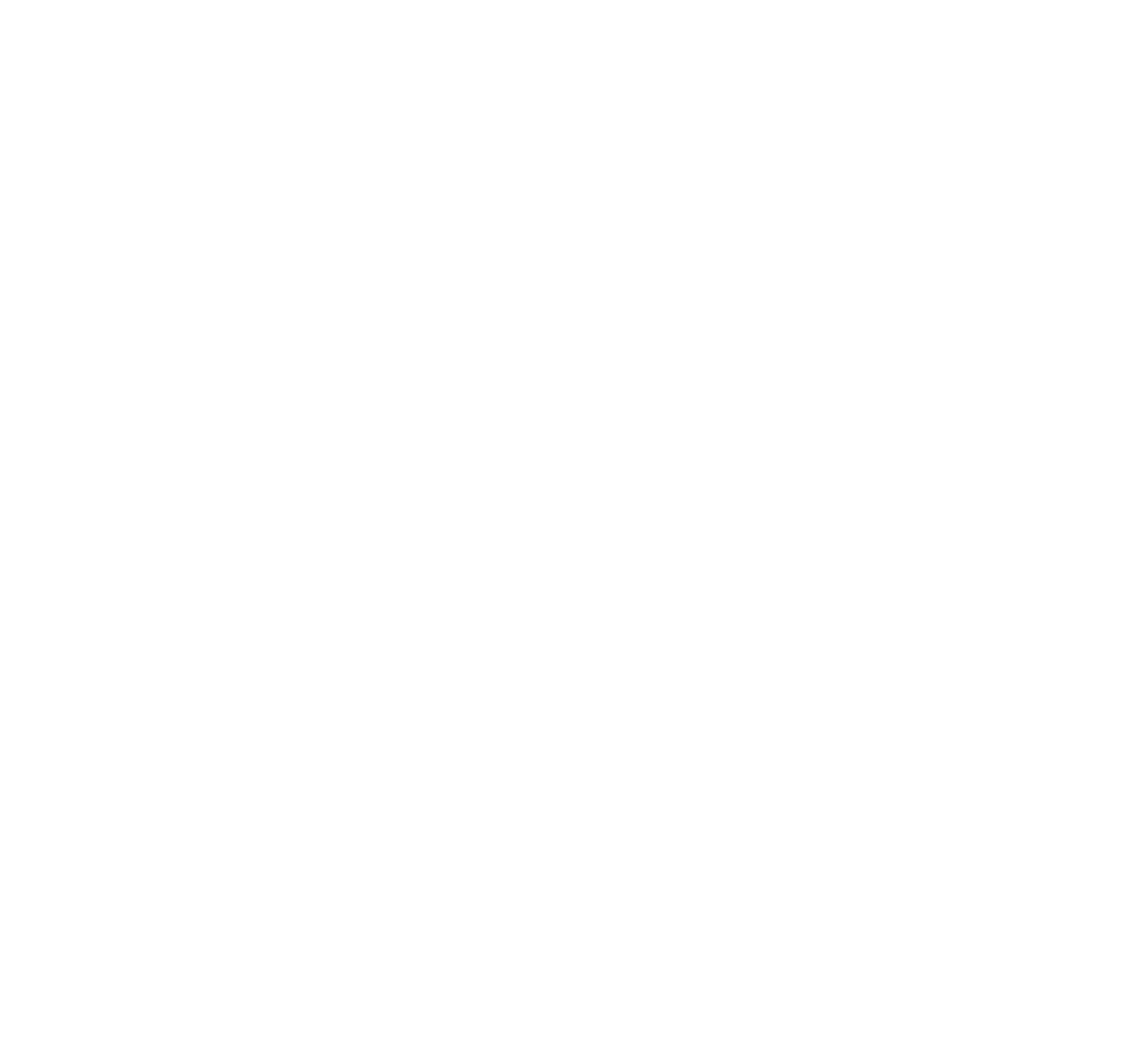}}
    \vspace*{8pt}
\caption{(Color online) Global equilibrium distribution functions $f_i^{eq}$ for volume fraction $\phi=0.8722$: (a) The equilibrium distribution function is obtained from the fitted mean-square displacement and Eq. (\ref{eq:msd_predicted}). There is a strong deviation between the measured and the theoretical data; (b) The equilibrium distribution function is calculated using the measured mean-square displacement from the MD simulation. The legend can be seen in Fig. \ref{fig:eq_distr_fn1} (omitted for clarity). }
\label{fig:eq_distr_fn2}
\end{figure}

To validate the analytical solution of the MDLG global equilibrium distribution function $f_i^{eq}$ in Eq. (\ref{eq:analytical_fi_eq_2}), we compare it to the measured $f_i^{eq}$ from the underlying MD simulation as a function of $a^2$  from Eq. (\ref{eq:analytical_fi_a2}). In Fig. \ref{fig:eq_distr_fn1} and Fig. \ref{fig:eq_distr_fn2}, we show that the theoretical prediction of the MDLG global equilibrium distribution function $f_i^{eq}$ agrees well with the measured one from them MD simulations as long as the mean-square displacement can be approximated from the theory Eq. (\ref{eq:msd_analytic}) as depicted in Fig. \ref{fig:mean_sq_displ}. However, once this relation breaks, the measured global equilibrium distribution function also starts to deviate from the theoretically predicted values as shown in Fig. \ref{fig:eq_distr_fn2}. For systems with volume fraction around and above $\phi=0.8722$, the MDLG global equilibrium distribution function is not well approximated from the theoretical mean-square displacement given in Eq. (\ref{eq:msd_predicted}).

As a remedy, instead of calculating the theoretical value of $\langle(\delta x)^2\rangle$, we can measure the mean-square displacement as a function of the time step $\Delta t$ directly and use this numerical value to obtain $a^2$. There is a noticeable difference between the MDLG global equilibrium distribution function obtained from the predicted and from the measured mean-square displacement as depicted in Fig. \ref{fig:eq_distr_fn2}(a) and Fig. \ref{fig:eq_distr_fn2}(b), respectively. Using the measured mean-square displacement, however, we recover the correct MDLG global equilibrium distribution function from theory even for high volume fractions.

We find, therefore, that the analytical prediction of the equilibrium distribution function is correct for a large range of volume fractions, even after the relation between the velocity correlation function and the mean-square displacement breaks and correlations become important. In this case, we need to use the actual mean-square displacement measured from the MD simulation instead of using the theoretical solution in Eq. (\ref{eq:velocity_corr_fn}) to obtain $a^2$.

\section{Conclusions}
\label{sec:conclusions}
In this work, we have investigated the behavior of the MDLG equilibrium function in respect to different volume fractions. We have shown that for a very large range of volume fractions, the equilibrium distribution function can be obtained from a single parameter $\lambda$ based on the relation between the velocity correlation function and the mean-square displacement of the underlying MD simulation. However, for higher volume fractions, this correlation does not apply, and the exponential fit from the velocity correlation function does not reproduce the right behavior of the mean-square displacement anymore, which leads also to wrong estimation of the equilibrium distribution function in respect to the measured one from the MD simulation. We found that by using the measured mean-square displacement instead of the theoretical, we can still recover the correct analytical MDLG equilibrium distribution function. We conclude that the MDLG equilibrium distribution function appears to be universally defined as a function of the mean-square displacement and the lattice size of the underlying MD system.

An exact knowledge of the equilibrium distribution is important, if one wishes to examine the deviation of a system from local equilibrium. In a future publication, we intend to investigate whether the MDLG collision operator can be expressed in terms of a BGK collision operator
$\Omega_i = \sum_i \Lambda_{ij}(f_j^{eq} - f_j)$
 with $\Lambda_{ij}$ being the relaxation matrix. Here the knowledge of an exact local equilibrium distribution is of the essence. Preliminary studies show that for non-equilibrium cases, we observe a small difference (approx. $0.5\%$) between the measured and the analytical global equilibrium distribution function. The reason for this discrepancy might be due to the assumption made in \cite{Parsa.2017}, that the probability distribution for the displacement is described as a Gaussian for all scales (from ballistic to diffusive regime). This assumption was made, because the probability distribution for short scales as well as the one for long scales can be described by a Gaussian distribution. However, intermediate probabilities between the ballistic and the diffusive regime could have different behavior, as for example the well-known \textit{cage effect} described by Chong et al.\cite{Chong.2002}. Those intermediate regimes due to their nature, might be better approximated, for example, by a combination of two Gaussian distributions. These considerations are outside the scope of the current paper and will be a subject of future research. 

Additionally, the knowledge of the equilibrium distributions is essential when analyzing the fluctuation behaviour of MDLG systems, which will elucidate the correct form of fluctuations for non-ideal systems. This is the subject of a following publication\cite{Parsa.2019}.

\section*{Acknowledgments}
AP acknowledges support from the German Federal Ministry of Education and Research (BMBF) in the scope of the project “Aerotherm” (reference numbers: 01IS16016A-B).
\label{sec:Acknowledgments}


\end{document}